\documentclass[aip,jcp,amsmath,amssymb,reprint]{revtex4-2}
\usepackage{graphicx, float, xcolor, pgf}
\usepackage{physics, siunitx}
\usepackage[caption=false, justification=centerlast]{subfig}

\makeatletter
\let\oldtheequation\theequation
\renewcommand\tagform@[1]{\maketag@@@{\ignorespaces#1\unskip\@@italiccorr}}
\renewcommand\theequation{(\oldtheequation)}
\let\cite\cite
\makeatother

\usepackage{hyperref}
\hypersetup{
  pdfauthor={Amartya Bose},
  pdftitle={A Multisite Decomposition of the Tensor Network Path Integral},
  colorlinks=true,
  allcolors=.
}

\begin{document}
\title{A Multisite Decomposition of the Tensor Network Path Integrals}
\author{Amartya Bose}
\thanks{Author to whom correspondence should be addressed}
\email{amartyab@princeton.edu}
\affiliation{Department of Chemistry, Princeton University, Princeton, New Jersey 08544}
\author{Peter L. Walters}
\email{peter.l.walters2@gmail.com}
\affiliation{Department of Chemistry, University of California, Berkeley, California 94720}
\affiliation{Miller Institute for Basic Research in Science, University of
  California Berkeley, Berkeley, California 94720}
\allowdisplaybreaks

\begin{abstract}
  Tensor network decompositions of path integrals for simulating open quantum
  systems have recently been proven to be useful. However, these methods scale
  exponentially with the system size. This makes it challenging to simulate the
  non-equilibrium dynamics of extended quantum systems coupled with local
  dissipative environments. In this work, we extend the tensor network path
  integral (TNPI) framework to efficiently simulate such extended systems. The
  Feynman-Vernon influence functional is a popular approach used to account for
  the effect of environments on the dynamics of the system. In order to
  facilitate the incorporation of the influence functional into a multisite
  framework (MS-TNPI), we combine a matrix product state (MPS) decomposition of
  the reduced density tensor of the system along the sites with a corresponding
  tensor network representation of the time axis to construct an efficient 2D
  tensor network. The 2D MS-TNPI network, when contracted, yields the
  time-dependent reduced density tensor of the extended system as an MPS. The
  algorithm presented is independent of the system Hamiltonian. We outline an
  iteration scheme to take the simulation beyond the non-Markovian memory
  introduced by solvents. Applications to spin chains coupled to local harmonic
  baths are presented; we consider the Ising, XXZ and the Heisenberg models,
  demonstrating that the presence of local environments can often dissipate the
  entanglement between the sites. We discuss three factors causing the system to
  transition from a coherent oscillatory dynamics to a fully incoherent
  dynamics. The MS-TNPI method is useful for studying a variety of extended
  quantum systems coupled with solvents.
\end{abstract}
\maketitle

\section{Introduction}

Quantum effects in dynamics are very important for studying charge or exciton
energy transfer in long chains and in understanding decoherence in systems of
qubits. To curtail the exponential growth of computational complexity, a
system-solvent description is often used. While in many cases, it is indeed
possible to limit the quantum description to only a small subspace of degrees of
freedom, for extended systems, however, this quantum subspace or ``system'' can
be quite large. Thus, the effectiveness of a typical system-solvent
decomposition might be compromised. Methods like density matrix renormalization
group~\cite{whiteDensityMatrixFormulation1992,schollwockDensitymatrixRenormalizationGroup2005,schollwockDensitymatrixRenormalizationGroup2011a,schollwockDensitymatrixRenormalizationGroup2011}
(DMRG) and its time-dependent
variant~\cite{whiteRealTimeEvolutionUsing2004,paeckelTimeevolutionMethodsMatrixproduct2019,xieTimedependentDensityMatrix2019}
(tDMRG) are very useful in simulating these large systems by decomposing the
wave function along the ``system'' axis using sequential singular value
decompositions (SVD). Multiconfiguration time-dependent Hartree (MCTDH) and its
multi-level version (ML-MCTDH) constitute another family of tensor network-based
algorithms that have also been commonly used to simulate non-equilibrium
dynamics. However, when vibrational or phononic modes are present, the wave
function-based nature of these algorithms pose significant computational
challenges.

Propagating the reduced density matrix is a lucrative option for simulating open
quantum systems. Path integrals based on the Feynman-Vernon influence functional
(IF)~\cite{feynmanTheoryGeneralQuantum1963} and the hierarchical equations of
motion (HEOM)~\cite{tanimuraTimeEvolutionQuantum1989} are rigorous methods for
incorporating the interactions between systems and solvents without having to
simulate the environmental degrees of freedom explicitly. While HEOM is, in
principle, exact for systems interacting with arbitrary harmonic baths,
practically, it has been mostly restricted to simulating the case of baths
described by Drude spectral densities. Attempts have been made to develop
efficient HEOM-based algorithms that are applicable to general spectral
densities.~\cite{tanimuraTimeEvolutionQuantum1989,duanStudyExtendedHierarchy2017,ikedaGeneralizationHierarchicalEquations2020,popescuChebyshevExpansionApplied2016,tanimuraNumericallyExactApproach2020,shiEfficientPropagationHierarchical2018,yanNewMethodImprove2020,yanEfficientPropagationHierarchical2021}
For cases of fermionic baths, it is also possible to simulate the influence
functional in an exact
manner.~\cite{segalNumericallyExactPathintegral2010,siminePathintegralSimulationsFermionic2013}
However, when the solvent is neither harmonic nor fermionic, but is
atomistically defined, the IF does not have a closed-form expression. Classical
trajectories are often used for estimating the influence functional for such
problems.~\cite{banerjeeQuantumClassicalPathIntegral2013,lambertQuantumclassicalPathIntegral2012,lambertQuantumclassicalPathIntegral2012a,waltersQuantumClassicalPath2015,waltersIterativeQuantumclassicalPath2016}
The quasi-adiabatic propagator path integrals
(QuAPI)~\cite{makriTensorPropagatorIterative1995a,makriTensorPropagatorIterative1995}
and related
methods~\cite{makriBlipDecompositionPath2014,makriBlipsummedQuantumClassical2016,makriIterativeBlipsummedPath2017}
are useful when simulating systems bilinearly coupled to harmonic baths.
Recently, tensor networks have also been shown to be useful in making
calculations with influence functionals more
efficient.~\cite{strathearnEfficientNonMarkovianQuantum2018,jorgensenExploitingCausalTensor2019,gribbenExactQuantumDynamics2020,yeConstructingTensorNetwork2021,boseTensorNetworkRepresentation2021,bosePairwiseConnectedTensor2021}

Thermal dynamics of extended systems coupled to a vibrational manifold poses
unique challenges to simulations. Even though there have been attempts to
incorporate these baths in terms of basis
sets,~\cite{renTimeDependentDensityMatrix2018} as mentioned before, wave
function-based methods like DMRG or tDMRG are typically not well-suited for
these problems. The computational complexity tends to grow because of the
entanglement between the system states and the bath modes.  Recently, the
modular path integral (MPI) approach has been
developed~\cite{makriCommunicationModularPath2018,makriModularPathIntegral2018,kunduModularPathIntegral2019,kunduModularPathIntegral2020,makriSmallMatrixModular2021}
and used to
study\cite{kunduRealTimePathIntegral2020,kunduExcitonVibrationDynamics2021}
extended systems with short-ranged interactions. Though powerful, MPI suffers
from an intrinsic inability to simulate the full density operator corresponding
to the extended system and a difficulty in dealing with entangled initial
states. This is because MPI treats the system sites sequentially. Based on
tensor network decompositions,~\citet{leroseInfluenceMatrixApproach2021} have
developed a method for simulating influence functionals for cases where the
system and the environment are made of indistinguishable particles. In
particular, they applied their method to calculate the reduced density matrix
for a specific site in a spin-chain.

Tensor network path integrals (TNPI)~\cite{boseTensorNetworkRepresentation2021}
offer an approach to influence functional-based path integral simulations that
can, quite naturally, be extended to handle problems involving extended systems
interacting with localized solvents. TNPI typically involves a matrix product
(MP) representation of the ``augmented propagator'' (AP). While TNPI offers
substantial gains in terms of compression of the path integral, it still does
not handle extended systems well. It is natural to wonder if the entire
time-evolving extended system can be represented in a compact form. Here, we
extend this TNPI representation to account for multiple system sites (or
particles) leading to a multisite version (MS-TNPI). The resulting
two-dimensional tensor network can be efficiently contracted and used to
simulate the reduced dynamics (in terms of an MP representation) of the extended
system. From a different perspective, MS-TNPI can be thought of as an extension
of tDMRG that incorporates Feynman-Vernon influence functionals to account for
interactions of sites with local solvents. Thus, it would be expected to be able
leverage the efficiency of the family of tensor network methods in adequately
representating the system.

As we will show, the MS-TNPI framework is independent of the structure of the
system Hamiltonian. All it requires is a matrix product operator (MPO)
representation of the propagator and a matrix product state (MPS) representation
of the system's initial density matrix. Thus, long-ranged interactions can be
accounted for without any change to the fundamental structure of the framework.
The time-evolved reduced density tensor corresponding to the entire extended
system is directly evaluated in the form of an MPS. The method is implemented
using the open-source ITensor library.~\cite{fishmanITensorSoftwareLibrary2020}
In Sec.~\ref{sec:method}, we develop the structure of MS-TNPI. The method is
illustrated with some examples in Sec.~\ref{sec:result}. We demonstrate the
ability of the method to compress the representation of large quantum systems.
Additionally, we explore various models of spin-chains and study the causes of
dissipation. Finally, we end the paper with some concluding remarks and future
prospects in Sec.~\ref{sec:conclusion}.

\section{Methodology}\label{sec:method}

Consider a system consisting of $P$ particles or sites each with its local vibrational degrees of freedom:
\begin{align}
  \hat{H} & = \hat{H}_0 + \sum_{i=1}^P \hat{V}_i,\label{eq:sysH}
\end{align}
where $\hat{H}_0$ is the system Hamiltonian and $\hat{V}_i$ is the Hamiltonian
encoding the system-vibration interactions localized on the
$i$\textsuperscript{th} site.


The system-vibration interactions generally have anharmonic terms, but under
Gaussian response theory, the effect of the anharmonic vibrations can be
accounted for by an equivalent harmonic bath on the $i$\textsuperscript{th}
site:
\begin{align}
  \hat{V}_i & = \sum_{l=1}^{N_{\text{osc}}} \frac{p^2_{i,l}}{2m_{i,l}} + \frac{1}{2}m_{i,l}\omega_{i,l}^2 \left(x_{i,l} - \frac{c_{i,l} \hat{s}_i}{m_{i,l} \omega_{i,l}^2}\right)^2\label{eq:harmonicbath},
\end{align}
where $\omega_{i,l}$ and $c_{i,l}$ are the frequency and coupling of the
$l$\textsuperscript{th} mode of the $i$\textsuperscript{th} site, respectively.
Additionally, $\hat{s}_i$ is the system operator, associated with the
$i$\textsuperscript{th} site, that couples the site with its local vibrations.
The site-vibration interaction is characterized by a spectral
density:~\cite{caldeiraPathIntegralApproach1983,makriLinearResponseApproximation1999}
\begin{align}
  J(\omega) & = \frac{\pi}{2}\sum_l\frac{c_l^2}{m_l\omega_l}\delta(\omega-\omega_l).
\end{align}
In case the vibrations are defined by atomistic Hamiltonians, it is often
possible to obtain the spectral density as a Fourier transform of the energy-gap
autocorrelation function simulated using classical trajectories. In case of
interactions with phonons, typically the modes can be exactly described by
harmonic oscillators, even without invoking Gaussian response theory.


The reduced dynamics of a system coupled to a harmonic bath is given by
\begin{align}
  \tilde{\rho}\left(S_{N}^\pm, N\Delta t\right) & = \Tr_{\text{bath}} \mel{S_{N}^+}{\tilde{\rho}(N\Delta t)}{S_{N}^-}\nonumber                        \\
                                                & = \sum_{S_0^\pm} \tilde{\rho}\left(S^\pm_0, 0\right) G\left(S_{0}^\pm, S_{N}^\pm, N\Delta t\right),
\end{align}
where $\tilde{\rho}$ is the system's reduced density tensor and $G$ is the AP.
In this notation, $S^\pm_n$ represents the forward-backward state of all the
sites at the $n$\textsuperscript{th} time point, with the state of the
$i$\textsuperscript{th} site at this time point being denoted by $s^\pm_{i,n}$.
In the absence of any coupling between the system and the bath, the
bare AP is given by:
\begin{align}
   & G^{(0)}\left(S_{0}^\pm, S_{N}^\pm, N\Delta t\right) = \sum_{S_{1}^\pm}\cdots\sum_{S_{N-1}^\pm} P_{S_{0}^\pm \cdots S_{N}^\pm}^{(0)}, \label{eq:bare_prop}                  \\
  \text{with}\nonumber                                                                                                                                                          \\
   & P_{S_{0}^\pm \cdots S_{N}^\pm}^{(0)} = K\left(S_{0}^\pm, S_{1}^\pm, \Delta t\right)\times\cdots \times K\left(S_{N-1}^\pm, S_{N}^\pm, \Delta t\right).\label{eq:bare_path}
\end{align}
Here, $P_{S_0^\pm \cdots S_N^\pm}^{(0)}$ is the bare path amplitude tensor and
$K\left(S^\pm_n,S^\pm_{n+1},\Delta t\right)$ is the forward-backward propagator
connecting the bare system at the $n$\textsuperscript{th} time point to the
$(n+1)$\textsuperscript{th} time point. Assuming the system Hamiltonian is time
independent,
\begin{align}
  K\left(S^\pm_n,S^\pm_{n+1},\Delta t\right) & = \mel{S_{n+1}^+}{\exp\left(-\tfrac{i}{\hbar} \hat{H}_0 \Delta
  t\right)}{S_n^+}\nonumber                                                                                   \\
                                             & \times\mel{S_n^-}{\exp\left(\tfrac{i}{\hbar} \hat{H}_0 \Delta
    t\right)}{S_{n+1}^-}.
\end{align}
Of course, if the Hamiltonian is explicitly time-dependent, we can simply obtain
the bare system forward-backward propagator by directly by solving the
time-dependent Schr\"{o}dinger's equation. However, this detail does not have
any impact on the formalism being developed. The time-independent Hamiltonian is
discussed purely for notational simplicity. From Eqs. \ref{eq:bare_prop} and
\ref{eq:bare_path}, we see that without the bath, the bare augmented propagator,
$G^{(0)}$, and consequently, the reduced dynamics can be evaluated iteratively.
The bath's presence, however, introduces non-Markovian effects that prevent such
a straightforward evaluation of the dynamics.

The dynamics of a system coupled to a harmonic bath is well-described by the
formalism of Feynman-Vernon influence functionals (IF). In this formalism, the
AP is given by the following
equations:~\cite{makriTensorPropagatorIterative1995a,makriTensorPropagatorIterative1995}
\begin{align}
  G\left(S_{0}^\pm, S_{N}^\pm, N\Delta t\right) & = \sum_{S_{1}^\pm}\cdots\sum_{S_{N-1}^\pm} P_{S_{0}^\pm \cdots S_{N}^\pm}                                              \\ \label{eq:aug_prop}
                                                & = \sum_{S_{1}^\pm}\cdots\sum_{S_{N-1}^\pm} F\left[\left\{S_{n}^\pm\right\}\right] P_{S_{0}^\pm \cdots S_{N}^\pm}^{(0)}
\end{align}
where $P_{S_{0}^\pm \cdots S_{N}^\pm}$ is the path amplitude tensor and
$F\left[\left\{S_{n}^\pm\right\}\right]$ is the influence functional for the
given forward-backward system path. The path amplitude tensor, has
$\mathcal{O}\left(d^{2NP}\right)$ coefficients, where $d$ is the dimensionality of a
typical system site, $N$ is the number of time points and $P$ is the number of
sites. The number of coefficients grows exponentially with both the number of
sites as well as the number of time points. In this work, we aim to combat this
exponential growth, by factorizing this unmanageably large tensor into a network
of smaller ones.

\begin{figure}
  \includegraphics[scale=0.20]{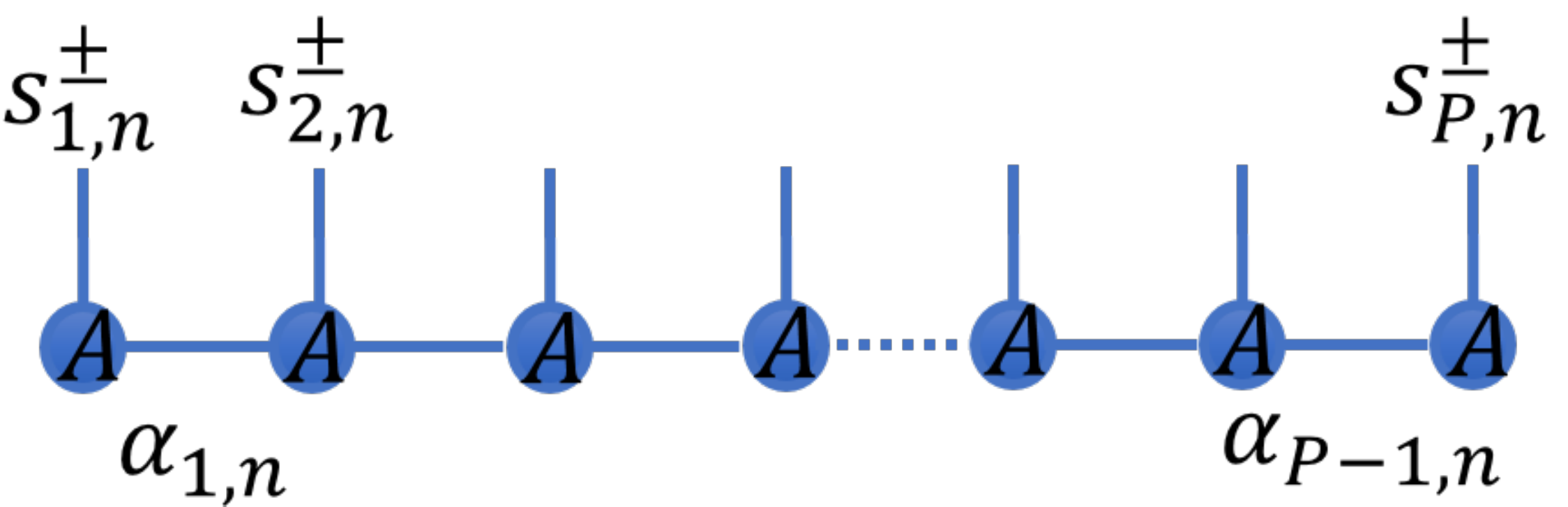}
  \caption{Matrix product state representation of the density matrix of the
  system as specified in Eq.~\ref{eq:Rho_MPS}.}\label{fig:Rho_MPS}
\end{figure}

For extended systems, it is often expected that the Hamiltonian's interactions
  and the correlations decrease as the separation between the sites increases.
  In these cases, the system can be very efficiently expressed as an MPS. With
  this in mind, we start by representing the system's reduced density matrix as
  an MPS
  \begin{align}
    \tilde{\rho}\left(S_{n}^\pm,n\Delta t\right)  = & \sum_{\left\{\alpha_{(i,n)}\right\}}
    A^{s_{1,n}^\pm}_{\alpha_{(1,n)}}  A^{s_{2,n}^\pm}_{\alpha_{(1,n)},\alpha_{(2,n)} }\nonumber                                                                         \\
                                                    & \cdots A^{s_{P-1,n}^\pm}_{\alpha_{(P-2,n)},\alpha_{(P-1,n)}}A^{s_{P,n}^\pm}_{\alpha_{(P-1,n)}}.\label{eq:Rho_MPS}
  \end{align}
  The indices that are used in the superscript are the ``site'' indices and they
  correspond to the forward-backward system state of the different particles.
  The indices in the subscripts, ${\left\{\alpha_{(i,n)}\right\}}$, are, in the
  DMRG literature, commonly called ``bond'' indices. In this work, however, we
  will refer to them as the ``spatial bond'' indices because this decomposition
  is done along the system or spatial dimension. Additionally, it should be
  noted that the parentheses associated with these indices are included solely
  for visual clarity. Both the site and spatial bond indices have two
  subscripts: the first subscript is for the site (or particle) number and the
  second one indicates the time point. This structure is visually depicted in
  Fig.~\ref{fig:Rho_MPS}. The dimensionality (or size) of the indices
  corresponding to the spatial bonds is closely related to the entanglement of
  the system. The maximum and average spatial bond dimension of the reduced
  density MPS after the $n$\textsuperscript{th} timestep is
  $m_{\rho}(n)=\max_{i} \left(\dim(\alpha_{(i,n)})\right)$ and
  $\bar{m}_{\rho}(n)=\tfrac{1}{P}\sum_i \dim(\alpha_{(i,n)})$, respectively.

In principle, this MPS factorization is exact; however, for an arbitrary density
  matrix, the maximum spatial bond dimension is $\mathcal{O}\left(d^{P}\right)$.
  In practice, these bond dimensions are truncated, with the maximum retained
  dimensionality of each being treated as a convergence parameter. In many
  cases, it is possible to accurately represent the density matrix with spatial
  bond dimensions significantly smaller than the theoretical maximum. In fact,
  if both the initial density matrix and system are separable (i.e., not
  entangled), the density matrix at all times is exactly represented with a
  maximum spatial bond dimension of one. In this case, the tensor,
  $A^{s_{i,n}^\pm}_{\alpha_{(i-1,n)},\alpha_{(i,n)}}$, is equivalent to the
  one-body reduced density matrix of the $i$\textsuperscript{th} particle after
  the $n$\textsuperscript{th} timestep. For cases where the Hamiltonian couples
  various system sites, the spatial bond dimensions are seen to grow with time,
  indicating an increase in the entanglement between the sites.

Next, we represent the forward-backward propagator, $K$, in the form of an MPO,
\begin{align}
  K\left(S_{n}^\pm, S_{n+1}^\pm, \Delta t\right)  = & \sum_{\left\{\alpha_{(i,n)}\right\}}
  W^{s_{1,n}^\pm, s_{1,n+1}^\pm}_{\alpha_{(1,n)}}  W^{s_{2,n}^\pm, s_{2,n+1}^\pm}_{\alpha_{(1,n)},\alpha_{(2,n)} }\nonumber                                                                              \\
                                                    & \cdots W^{s_{P-1,n}^\pm, s_{P-1,n+1}^\pm}_{\alpha_{(P-2,n)},\alpha_{(P-1,n)}}W^{s_{P,n}^\pm, s_{P,n+1}^\pm}_{\alpha_{(P-1,n)}}.\label{eq:prop_MPO}
\end{align}
As before, the superscripts correspond to the site indices and the subscripts
  correspond to the spatial bond indices. Unlike the reduced density MPS, which
  has a single site index ($s_{i,n}^\pm$) associated with each tensor, the
  forward-backward propagator MPO has two ($s_{i,n}^\pm$ and $s_{i,n+1}^\pm$).
  The two indices represent the same particle at different time points. Loosely
  speaking, the tensor, $W^{s_{i,n}^\pm,
  s_{i,n+1}^\pm}_{\alpha_{(i-1,n)},\alpha_{(i,n)}}$, can be thought of as an
  effective forward-backward propagator acting on the $i$\textsuperscript{th}
  particle.

Generally, any forward-backward propagator can be represented as an MPO, but
doing so may be as costly as directly solving the Schr\"{o}dinger equation.
Fortunately, using matrix product representations for studying dynamics of the
bare system has been a topic of intense research over the years. Various methods
like time-evolving block decimation
(TEBD)~\cite{daleyTimedependentDensitymatrixRenormalizationgroup2004,vidalEfficientSimulationOneDimensional07,whiteRealTimeEvolutionUsing2004}
and time-dependent DMRG
(tDMRG)~\cite{whiteRealTimeEvolutionUsing2004,xieTimedependentDensityMatrix2019}
have been developed to directly simulate the time evolution of wave functions of
extended systems with short-ranged interactions. For systems with long-ranged
interactions, the recently introduced MPO W\textsuperscript{I,II}
method~\cite{zaletelTimeevolvingMatrixProduct04} can generate very efficient
representations of the propagator. Additionally, a time-dependent variatonal
principle
(TDVP)~\cite{haegemanTimeDependentVariationalPrinciple2011,orusPracticalIntroductionTensor2014,yangTimedependentVariationalPrinciple2020}
approach has also been developed that allows for the treatment of arbitrary
Hamiltonians. While TEBD and MPO W\textsuperscript{I,II} calculate the
propagators, Krylov subspace-based methods and TDVP often approximate the action
of the propagator on the wave function. Detailed comparisons of these methods
for the purposes of simulating the propagator, especially in the context of the
current method, is extremely interesting and beyond the scope of this paper. For
the current development, we will simply assume that an MPO representation of the
forward-backward propagator is available.

To incorporate the Feynman-Vernon influence functional, in the standard TNPI
  framework,~\cite{boseTensorNetworkRepresentation2021,jorgensenExploitingCausalTensor2019,strathearnEfficientNonMarkovianQuantum2018,yeConstructingTensorNetwork2021}
  an SVD factorization is performed on the forward-backward system propagator
  along the time dimension, enabling us to represent the path amplitude tensor
  as an MPS:
  \begin{align}
    P_{S_{0}^{\pm}\cdots S_{N}^{\pm}} & = \sum_{\left\{\beta_{n}\right\}} T^{S_{0}^{\pm}}_{\beta_{0}}
    \cdots T^{S_{n}^{\pm}}_{\beta_{n-1}, \beta_{n}}
    \cdots T_{\beta_{N-1}}^{S_{N}^{\pm}}\label{eq:PA_MPS}
  \end{align}
  which can be acted upon by the influence functional MPO. Unlike in the MPS
  decomposition of the density matrix, Eq.~\ref{eq:Rho_MPS}, the site indices,
  here, correspond to the forward-backward state of the entire extended system
  at different time points. The bond indices, $\left\{\beta_n\right\}$,
  represent the coupling between different time points and will be referred to
  as the temporal bonds for the remainder of this work. More specifically, the
  dimensionality of the temporal bond indices is related to the length of the
  non-Markovian memory. In light of this MP factorization of the system
  described above, Eq.~\ref{eq:PA_MPS} can be rewritten as
  \begin{align}
    P_{S_{0}^{\pm}\cdots S_{N}^{\pm}} & = \sum_{\left\{\beta_{n}\right\}} \mathbb{T}^{S_{0}^{\pm}}_{\beta_{0}}
    \cdots \mathbb{T}^{S_{n}^{\pm}}_{\beta_{n-1}, \beta_{n}}
    \cdots\mathbb{T}_{\beta_{N-1}}^{S_{N}^{\pm}},\label{eq:MS-PA_MPS}
  \end{align}
  where the symbol $\mathbb{T}$ is an MP representation of the corresponding $T$
  tensors decomposed along the system or spatial dimension. Thus, we have a
  product of tensor products, i.e. a 2D array of tensors.

\begin{figure}
  \includegraphics[scale=0.20]{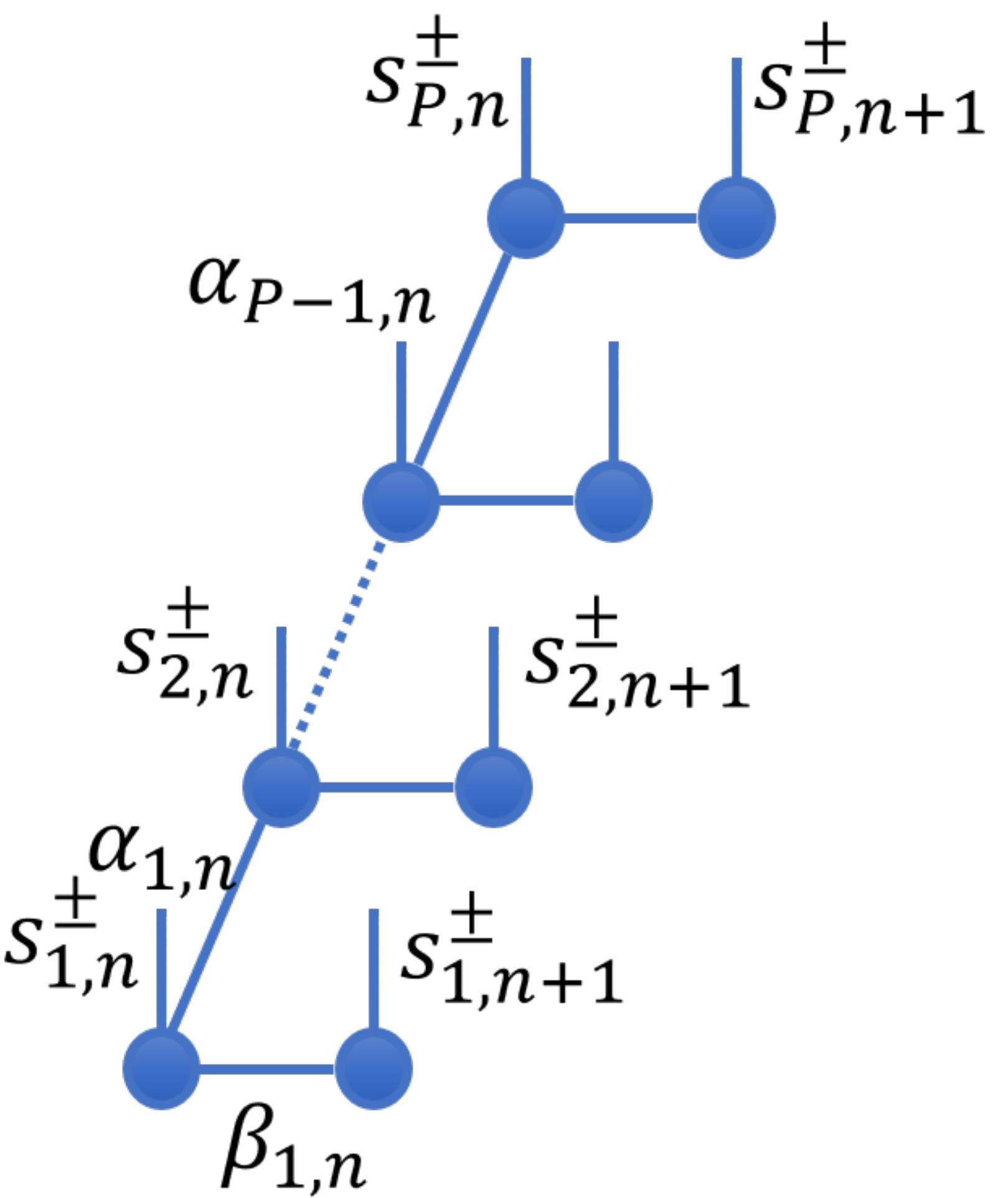}
  \caption{Factorization of the forward-backward MPO following
    Eqs.~\ref{eq:fbmpo1}--\ref{eq:fbmpo3}}\label{fig:prop_mpo}
\end{figure}

To facilitate this 2D decomposition of the path integral expression, we proceed
  by using SVD to factor the forward-backward propagator MPO in
  Eq.~\ref{eq:prop_MPO}:
\begin{align}
  W^{s_{1,n}^\pm, s_{1,n+1}^\pm}_{\alpha_{(1,n)}}                   & = \sum_{\beta_{(1,n)}} U^{s_{1,n}^\pm}_{\alpha_{(1,n)}, \beta_{(1,n)}} R^{s_{1,n+1}^\pm}_{\beta_{(1,n)}}                           \label{eq:fbmpo1}        \\
  W^{s_{i,n}^\pm, s_{i,n+1}^\pm}_{\alpha_{(i-1,n)},\alpha_{(i,n)} } & = \sum_{\beta_{(i,n)}} U^{s_{i,n}^\pm}_{\alpha_{(i-1,n)},\alpha_{(i,n)},\beta_{(i,n)}} R^{s_{i,n+1}^\pm}_{\beta_{(i,n)}},\medspace 1<i<P  \label{eq:fbmpo2} \\
  W^{s_{P,n}^\pm, s_{P,n+1}^\pm}_{\alpha_{(P-1,n)}}                 & = \sum_{\beta_{(P,n)}} U^{s_{P,n}^\pm}_{\alpha_{(P-1,n)}, \beta_{(P,n)}} R^{s_{P,n+1}^\pm}_{\beta_{(P,n)}}.\label{eq:fbmpo3}
\end{align}
where $U$ and $R$ are the factors obtained through the SVD procedure. The
  square-root of diagonal matrix of singular values has been absorbed into the
  $U$ and $R$ tensors. As per our convention, the bonds along the spatial and
  temporal dimensions are denoted by $\alpha$ and $\beta$, respectively.
  Figure~\ref{fig:prop_mpo} shows this structure in the form of a tensor
  diagram.

\begin{figure}
  \includegraphics[scale=0.20]{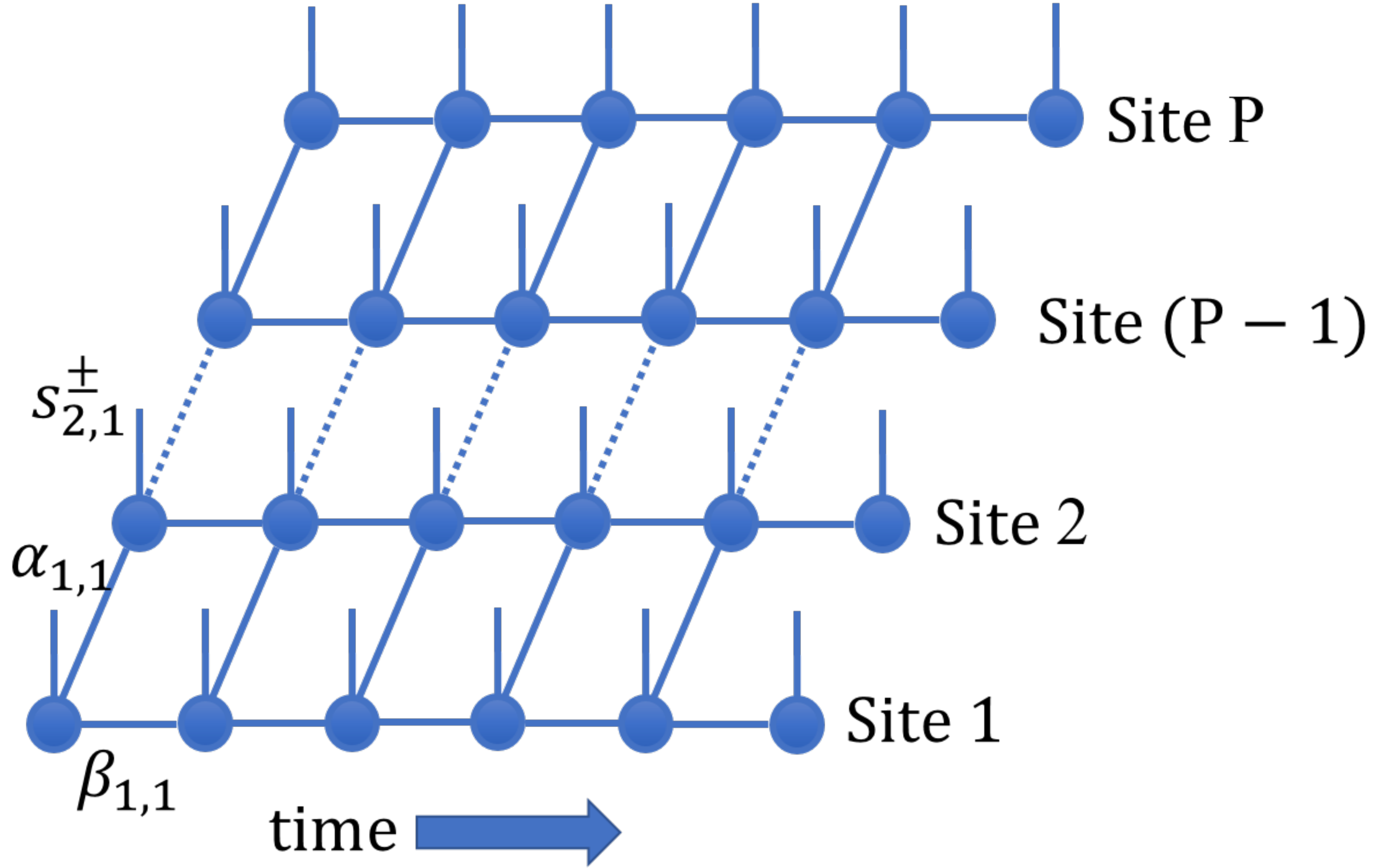}
  \caption{Factorization of the forward-backward MPO.}\label{fig:mattress_diag}
\end{figure}

Now we can put the expressions together to derive the tensors constituting the
  MPs, $\mathbb{T}$, in Eq.~\ref{eq:MS-PA_MPS}. Generally speaking, each of
  these constituent tensors, represented here by $M$, possesses five indices:
  one site, $s^\pm_{i,n}$, and four bonds ($\alpha_{(i,n)}$, $\beta_{(i,n)}$,
  $\alpha_{(i-1,n)}$ and $\beta_{(i,n-1)}$), where the values of $i$ and $n$
  correspond to the location of the tensor in the 2D grid structure, which is
  illustrated in Fig.~\ref{fig:mattress_diag}. It's worth noting that the
  tensors on the edges of the grid have a slightly different structure, as the
  number of bond indices differ. The tensors corresponding to the initial time
  point, or equivalently the first column, are given as:
\begin{align}
  M^{s^\pm_{1,0}}_{\alpha_{(1,0)}, \beta_{(1,0)}}                   & = U^{s^\pm_{1,0}}_{\alpha_{(1,0)}, \beta_{(1,0)}}                   \\
  M^{s^\pm_{i,0}}_{\alpha_{(i,0)}, \beta_{(i,0)}, \alpha_{(i-1,0)}} & = U^{s^\pm_{i,0}}_{\alpha_{(i-1,0)}, \alpha_{(i,0)}, \beta_{(i,0)}} \\
  M^{s^\pm_{P,0}}_{\alpha_{(P-1, 0)}, \beta_{(P,0)}}                & = U^{s^\pm_{P,0}}_{\alpha_{(P-1,0)}, \beta_{(P,0)}}.
\end{align}
The expressions for the final point, last column:
\begin{align}
  M^{s^\pm_{1,N}}_{\beta_{(1,N-1)}} & = R^{s^\pm_{1,N}}_{\beta_{(1,N-1)}}\label{eq:lastcol1}  \\
  M^{s^\pm_{i,N}}_{\beta_{(i,N-1)}} & = R^{s^\pm_{i,N}}_{\beta_{(i,N-1)}}\label{eq:lastcol2}  \\
  M^{s^\pm_{P,N}}_{\beta_{(P,N-1)}} & = R^{s^\pm_{P,N}}_{\beta_{(P,N-1)}}\label{eq:lastcol3}.
\end{align}
Lastly, for an intermediate time point, $n$:
\begin{align}
  M^{s^\pm_{1,n}}_{\alpha_{(1,n)}, \beta_{(1,n)}, \beta_{(1,n-1)}}                   & = R^{s^\pm_{1,n}}_{\beta_{(1,n-1)}} U^{s^\pm_{1,n}}_{\alpha_{(1,n)}, \beta_{(1,n)}}                   \\
  M^{s^\pm_{i,n}}_{\alpha_{(i,n)}, \beta_{(i,n)}, \alpha_{(i-1,n)}, \beta_{(i,n-1)}} & = R^{s^\pm_{i,n}}_{\beta_{(i,n-1)}} U^{s^\pm_{i,n}}_{\alpha_{(i-1,n)}, \alpha_{(i,n)}, \beta_{(i,n)}} \\
  M^{s^\pm_{P,n}}_{\beta_{(P,n)}, \alpha_{(P-1,n)}, \beta_{(P,n-1)}}                 & = R^{s^\pm_{P,n}}_{\beta_{(P,n-1)}} U^{s^\pm_{P,n}}_{\alpha_{(P-1,n)}, \beta_{(P,n)}}.
\end{align}
Notice that here, the sites on the final time
point are not connected together, Eqs.~\ref{eq:lastcol1}\,--\,\ref{eq:lastcol3},
inheriting the fundamental asymmetry between the initial and final time points
in the structure in Fig.~\ref{fig:prop_mpo}.

The flexibility of this factorization becomes apparent when the bath
interactions, in form of an IF, are incorporated. The 2D structure discussed
till now can be thought of as a series of generalized tensor products along the
``columns'' that represent the state of the full system at a given time point,
or along the ``rows'' that represent the state of one site at all times (i.e.
the path amplitude tensor of that particular site). While thinking of it as a
collection of columns manifestly connects the method to its tDMRG heritage, its
identification as a collection of ``row'' tensors serve to illustrate how this
multisite method is related to
AP-TNPI.~\cite{boseTensorNetworkRepresentation2021} (On a different note, in
this picture, MPI would be similar to a method that iterates over the rows of
this 2D structure.) If there was no interaction between the sites, the rows
would separate out and every site would behave like the standard TNPI method.
This makes it quite simple to account for the influence functional.

Because we are considering site-local baths, the total IF is just a product of
the IFs on each of the sites. The structure of the IF is the same irrespective
of the site. Hence, if we consider the path of the $i$\textsuperscript{th} site
as being given by $\left\{s^\pm_{i,n}\right\}$, then the
IF~\cite{feynmanTheoryGeneralQuantum1963} is:
\begin{align}
  F\left[\left\{s^\pm_{i,n}\right\}\right] & = \exp\left(-\frac{1}{\hbar}\sum_{0\le k\le N}\Delta s_{i,k}\sum_{0\le k'\le k} (\Re(\eta_{kk'})\Delta s_{i,k'}\right.\nonumber \\
                                           & \left.\vphantom{\sum_{0\le k\le N}}+2i\Im(\eta_{kk'})\bar{s}_{i,k'})\right)\label{eq:IF}
\end{align}
where $\Delta s_{i,k} = s^+_{i,k} - s^-_{i,k}$ and $\bar{s}_{i,k} =
  \tfrac{s^+_{i,k} + s^-_{i,k}}{2}$ and $\eta_{kk'}$ are the coefficients
  obtained by discretising the bath response function along the quasi-adiabatic
  path.~\cite{makriTensorPropagatorIterative1995a,makriTensorPropagatorIterative1995}
  It is possible to have different baths associated with different sites leading
  to a site-dependent $\eta$-coefficient and site-dependent influence
  functional, however for notational convenience, we describe the method
  assuming the baths on different sites are characterized by the same spectral
  density.

\begin{figure}
  \includegraphics[scale=0.20]{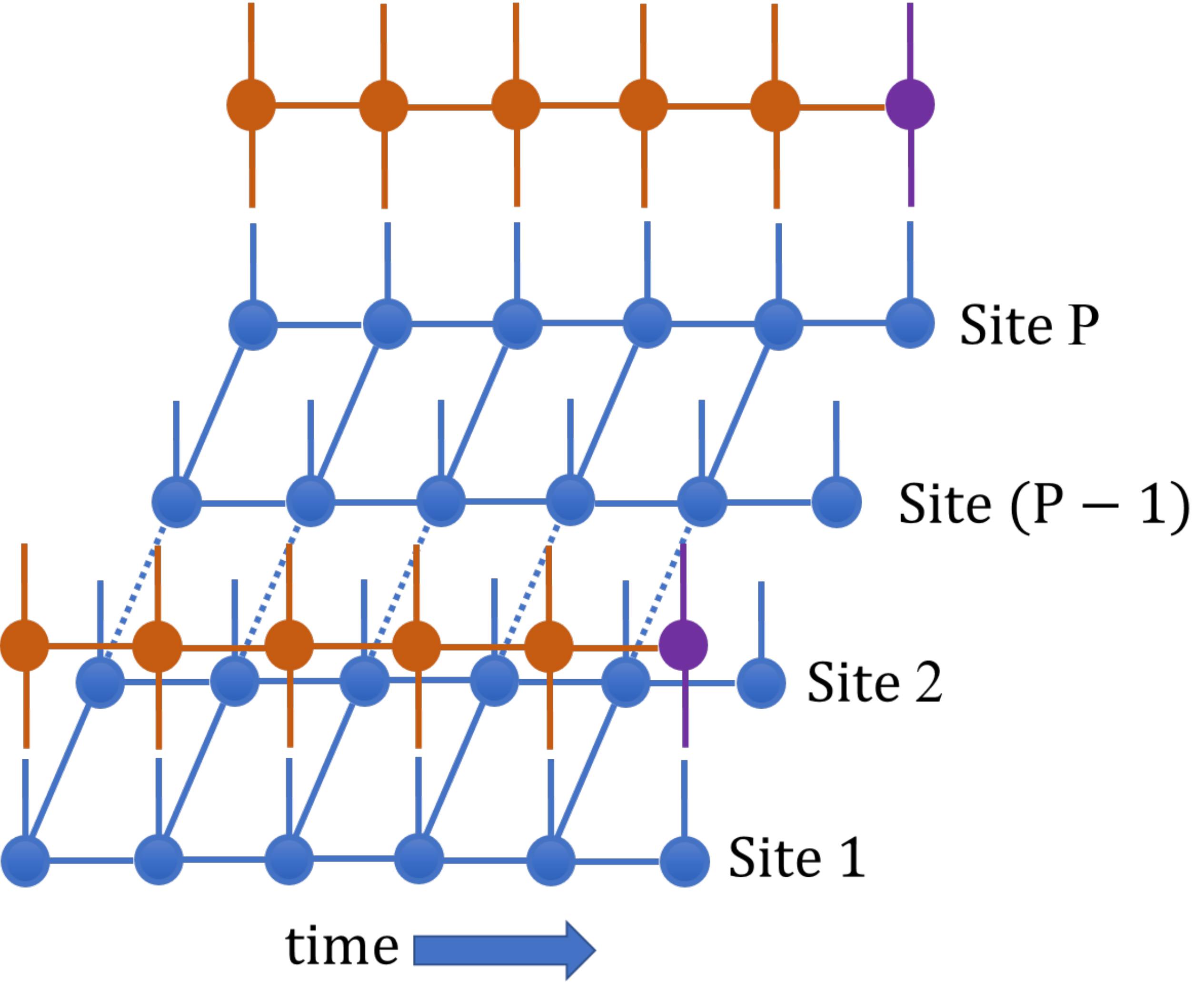}
  \caption{Schematic showing the application of the influence functional MPO to
    the 2D MS-TNPI structure. Only the influence functionals for the first and
    the last sites are shown here. Purple vertices correspond to the
    ``projection'' operator on the last time point (cf.
    Ref~\citep{boseTensorNetworkRepresentation2021}).}\label{fig:mattress_if}
\end{figure}

We have already discussed the analytical form for the matrix product operator
for the IF.~\cite{boseTensorNetworkRepresentation2021} Following that procedure,
Eq.~\ref{eq:IF} is factorized based on the $k$ time point:
\begin{align}
  F_k\left[\left\{s^\pm_{i,n}\right\}\right] & = \exp\left(-\frac{1}{\hbar}\Delta s_{i,k}\sum_{0\le k'\le k} (\Re(\eta_{kk'})\Delta s_{i,k'}\right.\nonumber \\
                                             & \left.\vphantom{\sum_{0\le k\le N}}+2i\Im(\eta_{kk'})\bar{s}_{i,k'})\right)
\end{align}
and each $F_k$ is given an MPO-representation, $\mathbb{F}_k$. The MPOs are
applied to each row of the 2D multisite TNPI structure for each system site in
order of increasing $k$ as detailed in
Ref.~\citep{boseTensorNetworkRepresentation2021}. The operations on the first
and the last rows (sites) are schematically indicated in
Fig.~\ref{fig:mattress_if}.

The resulting the tensor network corresponds to the path amplitude tensor. By
tracing over the internal system indices as well as contracting all the temporal
bonds, we can obtain the AP expressed as an MPO. This is achieved most easily by
considering the network as a sequence of columns, each representing the state of
the system at a given time point. Tracing over the internal system indices turns
these columns into MPOs. From here, the AP MPO is obtained by multiplying all
the column MPOs together in a sequential manner, thereby contracting the
temporal bonds. By applying this AP MPO to an MPS representing the initial state
of the system, we can obtain the corresponding time-evolved final state. This
AP-formulation is particularly helpful if we are interested in studying the
behaviors of a variety of different initial states. However, usually only the
dynamics arising from a specific initial system state is desired. In such cases,
we can obtain the resulting final state in a more efficient manner by reducing
the problem to one of a sequential application of the column MPOs to the initial
state MPS. As MPS-MPO operations are much cheaper than MPO-MPO operations, the
AP MPO should not be computed unless it is required. For the examples given
here, we will restrict our attention to the MPS-MPO contraction scheme.

\begin{figure}
  \includegraphics[scale=0.20]{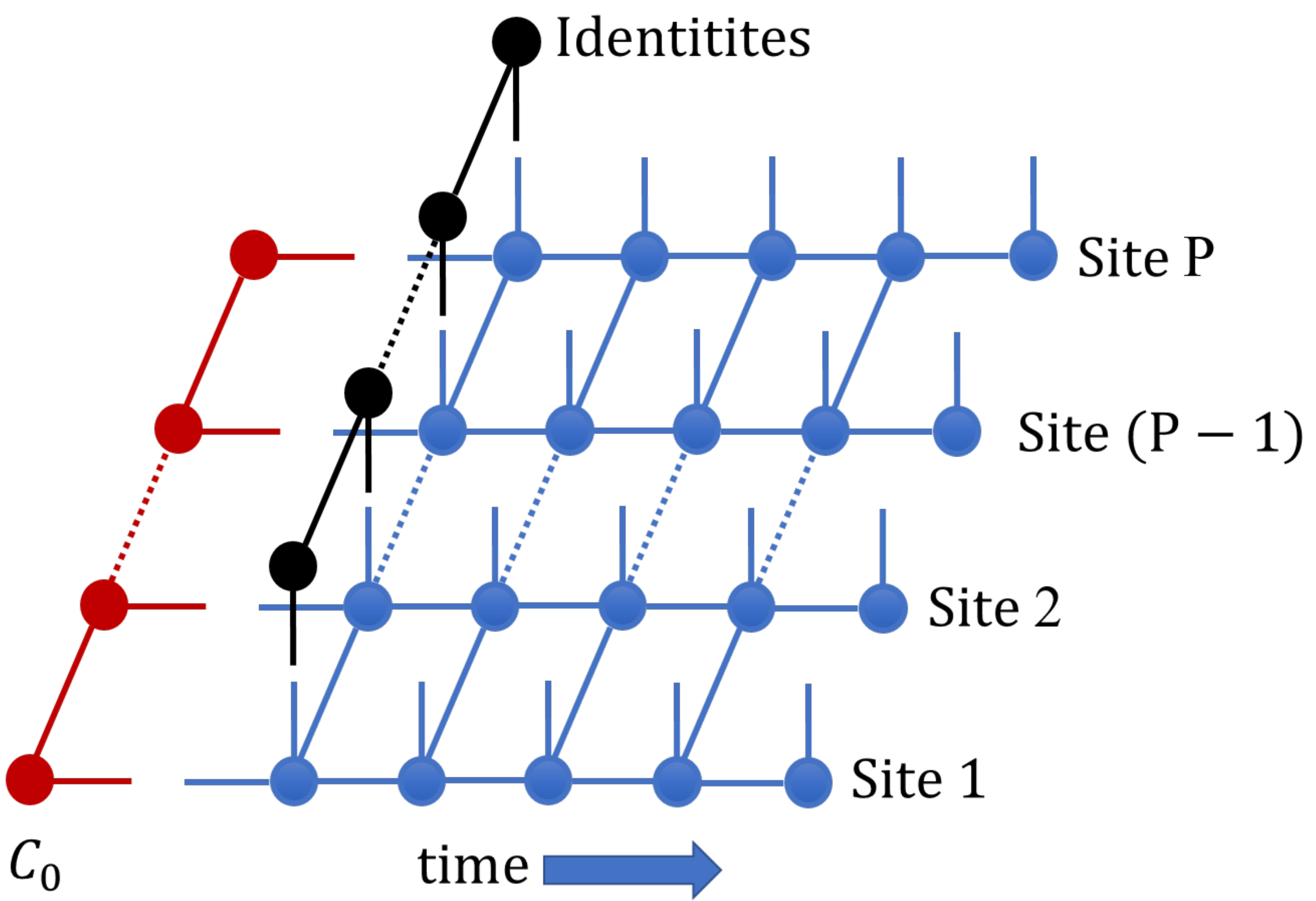}
  \caption{Iteration of the MS-TNPI network. At step (5), notice that the first
    column is not an MPO. After contraction with identity MPS, it becomes a
    MPO.}\label{fig:mattress_iter}
\end{figure}

It is well-known that for simulations in the condensed phase, the non-Markovian
memory does not extend for all of history. This means that the paths can be
truncated after $L$ time steps, to simulate a non-Markovian memory of time-span
$L\Delta t$. To develop such an iteration procedure, one needs to know the full
state of the system at any time point. We have access to that information for
the extended system in the form of the column corresponding to the relevant
time-point in the 2D lattice structure, Fig.~\ref{fig:mattress_diag}. Iteration
can also be done in two ways --- for the AP as done in
Ref.~\citep{boseTensorNetworkRepresentation2021}, or as typically done, for a
particular initial
state.~\cite{makriTensorPropagatorIterative1995,makriTensorPropagatorIterative1995a}
Once again in the interest of simplicity of discussion and computational
efficiency, we describe the iteration scheme for a particular initial reduced
density tensor, $C_0 = \tilde\rho(0)$. For a simulation with memory length $L$,
in the iterative regime, there would always be $L+1$ columns, $C_n$ for $1\le
n\le L+1$. The iterative procedure is outlined below. In the following steps for
iterative propagation, a tensor contraction is represented by $\otimes$.
\begin{enumerate}
  \item Update $C_0$ by applying the MPO, $C_1$ to it. $C_0\leftarrow C_1
          \otimes C_0$.
  \item Copy the other columns in by ``sliding them'' back by one.
        $C_n\leftarrow C_{n+1}$ for $n<L$.\\
        Notice at this stage that the new first column is no longer an MPO.
  \item Update the second last column, $C_L$ by multiplying by the corresponding
        $U$ tensors obtained from the SVD decomposition of the forward-backward
        MPO in Eqs.~\ref{eq:fbmpo1}--\ref{eq:fbmpo3}.
  \item Insert the last column according to
        Eqs.~\ref{eq:lastcol1}--\ref{eq:lastcol3}.
  \item Apply the IF MPO in a row-wise fashion.
  \item Trace over the site indices of $C_1$ to obtain an MPO in the first
        column.
\end{enumerate}
The steps 1 through 6 are repeated as many times as required. The salient ideas
of the iterative procedure are schematically outlined in
Fig.~\ref{fig:mattress_iter}.

Finally, given the immense difficulty of the problem at hand, it is of interest
to estimate the complexity of the algorithm outlined. The two most
computationally demanding operations that occur during each time step are the
application of the IF MPO and the contraction of the resulting tensor network.
Roughly speaking, the cost of applying the IF MPO is
$\mathcal{O}\left(m_{t}^3w_{p}^2w_{I}^2d^2\right)$ and cost of the contraction
is $\mathcal{O}\left(m^3w_{p}^2m_{t}\right)$. Here, $m_{t}$ is the maximum
temporal bond dimension, $m$ is the maximum bond dimension of the contracting
MPS, $w_{I}$ is the maximum bond dimension of the IF MPO. The typical bond
dimension of the forward-backward propagator of the bare system is denoted by
$w_{p}$ and $d$ is the dimensionality of a typical system site. Though the
magnitude of $m_t$ might be dependent on the memory length, $L$, the exponential
growth of complexity within memory is effectively
curtailed.~\cite{boseTensorNetworkRepresentation2021}
Since the local vibrational baths would typically consist of high frequency
modes (implying that the non-Markovian memory length, $L$, is not very large),
and the site-site couplings would be quite high, the cost would probably be
dominated by the contraction process. However, if the local baths are very
strongly coupled, the temporal bond dimension would grow much faster than the
bond dimension along the system axis, and the pattern would be reversed.

\section{Results}\label{sec:result}

For the purposes of illustrating the multisite TNPI method, we consider spin
chains with nearest-neighbor intersite coupling. The Hamiltonian is given as
\begin{align}
  \hat{H}_0 & = \sum_{i=1}^P \hat{h}^{(1)}_i + \sum_{i=1}^{P-1}  \hat{h}^{(2)}_{i,i+1}
\end{align}
where
\begin{align}
  \hat{h}^{(1)}_i & = \epsilon \hat{\sigma}_z^{(i)} - \hbar\Omega \hat{\sigma}_x^{(i)}\label{eq:onebody}
\end{align}
is the one-body term. The strength of the transverse field is $\hbar\Omega$, and
$\epsilon$ represents any asymmetry present in the system due to a longitudinal
field. The two-body interaction term is given by a general nearest-neighbor
Hamiltonian:
\begin{align}
  \hat{h}^{(2)}_{i,j} & = \delta_{j, i+1} \left(J_x \hat{\sigma}_x^{(i)}\hat{\sigma}_x^{(j)}
  + J_y\hat{\sigma}_y^{(i)}\hat{\sigma}_y^{(j)}
  + J_z\hat{\sigma}_z^{(i)}\hat{\sigma}_z^{(j)}\right).\label{eq:twobody}
\end{align}
Here, $\hat{\sigma}^{(i)}_x$, $\hat{\sigma}^{(i)}_y$, $\hat{\sigma}^{(i)}_z$ are
the Pauli spin matrices on the $i$\textsuperscript{th} site. Each of the sites
is also coupled with its vibrational degrees of freedom described by the
harmonic bath given in Eq.~\ref{eq:harmonicbath} with the system-bath coupling
operator $\hat{s}_i = \hat{\sigma}_z^{(i)}$. For the examples shown here, the
harmonic bath is characterized by an Ohmic spectral density with an exponential
decay:
\begin{align}
  J(\omega) & = \frac{\pi}{2}\hbar\xi\omega\exp\left(-\frac{\omega}{\omega_c}\right)\label{eq:Ohmic}
\end{align}
where $\xi$ is the dimensionless Kondo parameter, and $\omega_c$ is the
characteristic cutoff frequency. In Appendix~\ref{app:MPOprop}, we outline the
second-order Suzuki-Trotter splitting TEBD scheme used here to construct the
forward-backward propagator MPO.

Depending on the nature of the intersite coupling, there are many models for
interacting spin chains. Here, we consider the dynamics of the Ising model, the
XXZ model and the Heisenberg model with $P=31$ sites, coupled to site-local
harmonic baths. The states of each system site are labeled $\ket{+1}$ and
$\ket{-1}$, which are eigenstates of the $\hat{\sigma}_z$ operator with
eigenvalue of $+1$ and $-1$ respectively. Though the initial condition for
MS-TNPI can be any arbitrary reduced density MPS (for example, a DMRG ground
state), here for simplicity, it is defined as the direct product state of all
spins being in $\ket{+1}$. An SVD compression scheme with a truncation threshold
of $\chi$, which is treated as a convergence parameter, is applied to the
propagator MPO as well as the results of any MPS-MPO multiplications. Under this
scheme, singular values, $\lambda_{n}$, are discarded such that
\begin{align}
  \frac{\sum_{n\in\text{discarded}}\lambda_{n}^{2}}{\sum_{n}\lambda_{n}^{2}}<\chi.
\end{align}

So, for the following simulations, in addition to the $\Delta t$ and $L$
    parameters standard to path integral simulations, $\chi$ is also a
    convergence parameter. Unlike the single site version, in the most general
    multisite case, there are two factors that affect the Trotter error, and
    correspondingly the converged $\Delta t$. First, there is a Trotter error
    associated with the system-solvent splitting. This is very similar to the
    single-site case. Additionally, the multisite problem also has a Trotter
    error associated with the splitting of the bare system propagator under the
    TEBD scheme.

Transition from an underdamped coherent to a fully incoherent behavior is a
  hallmark of system-solvent dynamics. Here, we have three different parameters
  that affect the nature of dynamics --- the strength of the intersite coupling
  (quantifying the ability of the spin chain to act as a self-bath), the
  strength of the system-solvent coupling, and the temperature.


\begin{widetext}
  \begin{minipage}{\linewidth}
    \begin{figure}[H]
      \subfloat[Convergence with respect to $L$ for $\Omega\Delta t = 0.25$]{\includegraphics{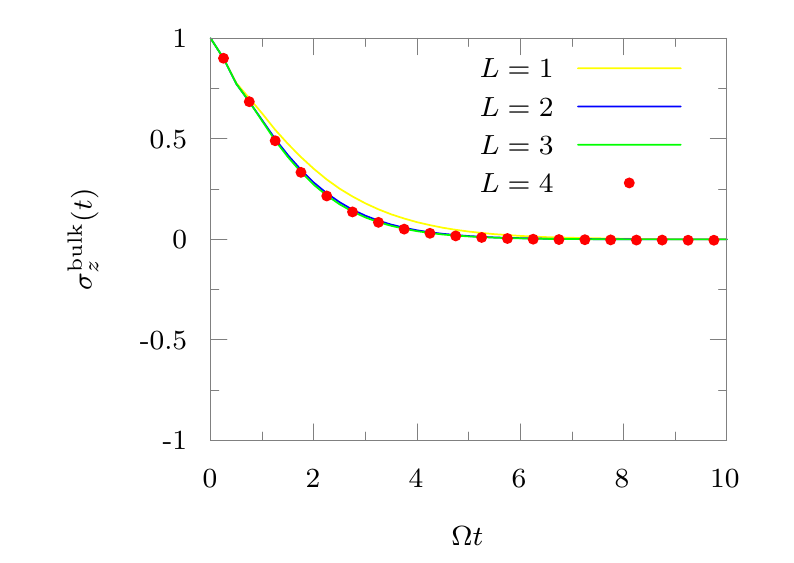}}
      ~ \subfloat[Convergence with respect to $\Delta t$ for $L \Delta t =
          1$]{\includegraphics{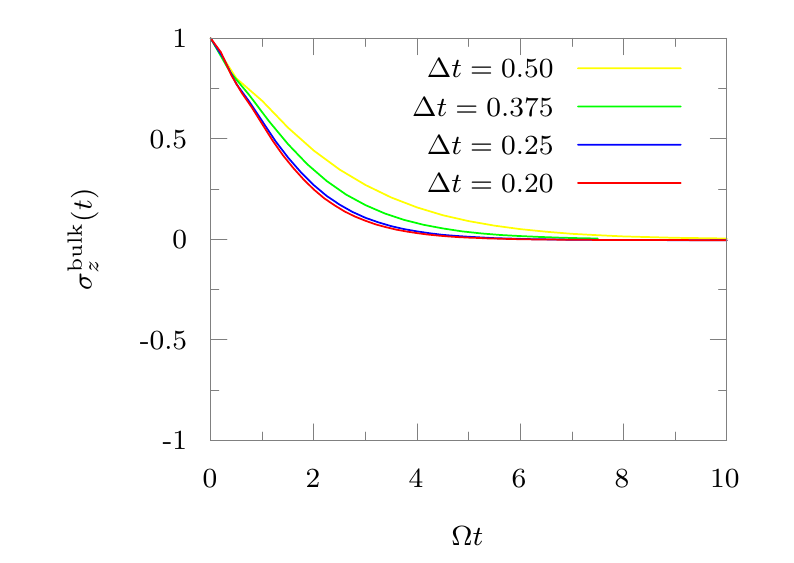}}

      \caption{Covergence of dynamics with
        respect to various parameters for the Ising model with $J_z =
          1.6$.}\label{fig:isingMem}
    \end{figure}
  \end{minipage}
\end{widetext}

\subsection{Ising Model}

We first consider a transverse-field Ising model ($J_x = J_y = 0$ and $J_z\ne
  0$) coupled with local vibrations. The longitudinal field is absent, $\epsilon
  = 0$, and a unit transverse field $\Omega = 1$ is applied.  The dynamics is
  simulated for different values of the intersite coupling, $J_z = \pm 0.2, \pm
  0.4, \pm 0.8$ and $\pm 1.6$. Here the bath is characterized by $\xi = 0.25$,
  $\omega_c = 5\Omega$, and it is held at an inverse temperature of
  $\hbar\beta\Omega = 1$. The convergence of $J_z = 1.6$ is the most difficult.
  Therefore, in Fig.~\ref{fig:isingMem}, we demonstrate the convergence patterns
  for this parameter.

Figure~\ref{fig:ising} shows $\expval{\hat{\sigma}_z(t)}$ for the
16\textsuperscript{th} spin in the chain. We observed that the finite size
effects of the chain were limited only to a few edge sites and the dynamics of
this middle monomer remained unaffected within the time-span of simulation,
implying that this is the bulk dynamics. A timestep of $\Omega\Delta t = 0.25$
was found to be converged for Fig.~\ref{fig:ising}~(a)--(c). When the bath was
present, a memory length of $L = 4$ was used, though acceptable convergence was
already achieved at $L = 3$. According to Fig.~\ref{fig:isingMem}, for $|J_z| =
1.6\hbar\Omega$ (Fig.~\ref{fig:ising}~(d)), $\Omega\Delta t$ was converged at
0.20 and $L$ at 5. A converged compression was done at a cutoff of
$\chi=10^{-11}$. The dynamics of the bare system is shown for the various cases
in dashed lines. It, unlike the dynamics in presence of the dissipative bath,
remains the same irrespective of the sign of $J_z$. For the bare dynamics, we
could use the MS-TNPI method and it would reduce to a density matrix version of
TEBD. However, for efficiency, we propagated the wave function using TEBD with
the same cutoff of $\chi = 10^{-11}$. In Fig.~\ref{fig:ising}, we see that
increasing the intersite couplings leads to a more incoherent dynamics. For the
high intersite coupling case, the excess dissipation happens primarily due to
the extended Ising chain.

The case of $J_z = -0.2\hbar\Omega$ (Fig.~\ref{fig:ising}~(a) red line) was
discussed by \citet{makriSmallMatrixModular2021} for a system with 10 sites. We
recover identical edge spin dynamics with our method. We observe that the finite
size of the chain affects more sites when $J_z$ is larger (not shown in figure).
This is because, for larger values of $J_z$, the sites ``know'' more about their
neighbors, making the difference between an edge site with only one neighbor and
a middle site with two neighbors more obvious. It is interesting that, though
the difference of sign in the values of $J_z$ does not impact the dynamics of
the bare system, it leads to profound differences once the bath is coupled.
Positive values of $J_z$ appear to make the population dissipation faster.

\begin{widetext}
  \begin{minipage}{\linewidth}
    \begin{figure}[H]
      \centering
      \subfloat[$|J_z| = 0.2\hbar\Omega$.]{\includegraphics{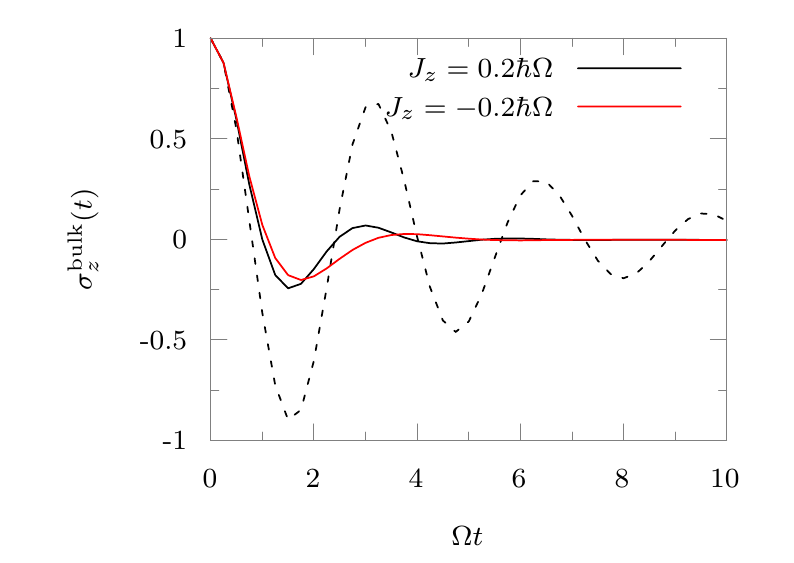}} ~\subfloat[$|J_z| =
          0.4\hbar\Omega$.]{\includegraphics{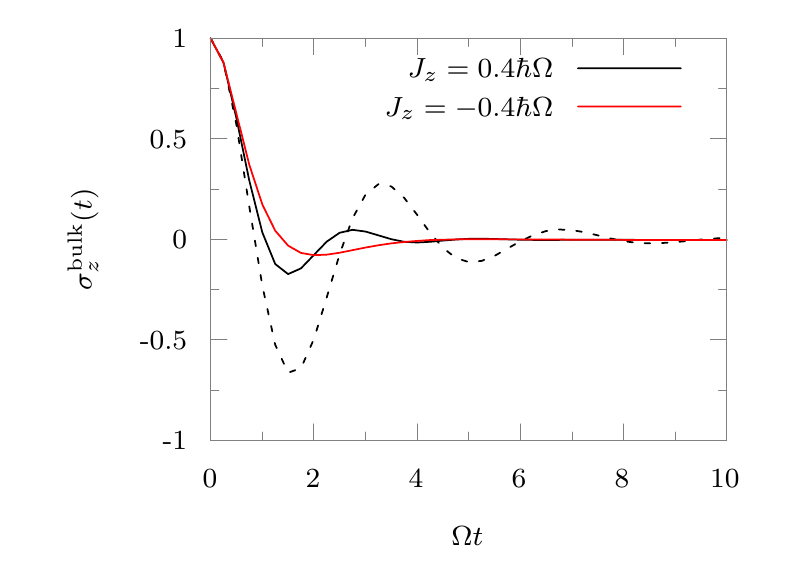}}

      \subfloat[$|J_z| = 0.8\hbar\Omega$.]{\includegraphics{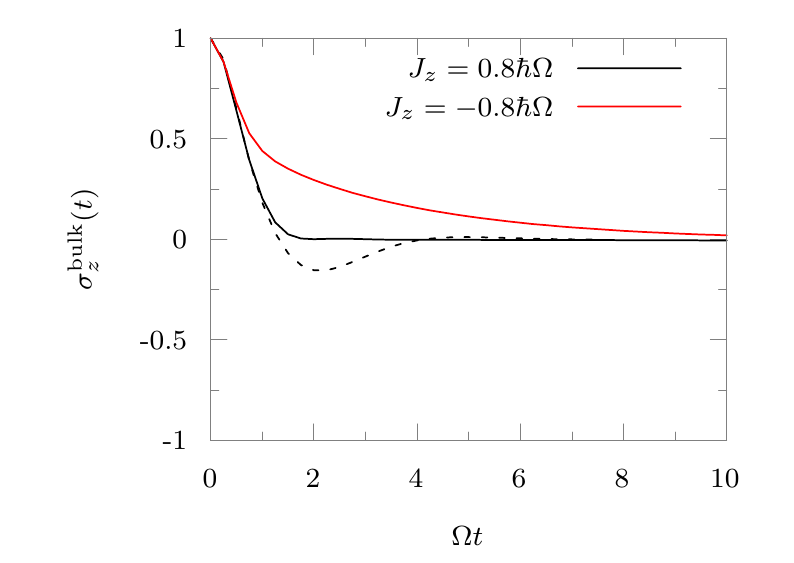}} ~\subfloat[$|J_z| =
          1.6\hbar\Omega$.]{\includegraphics{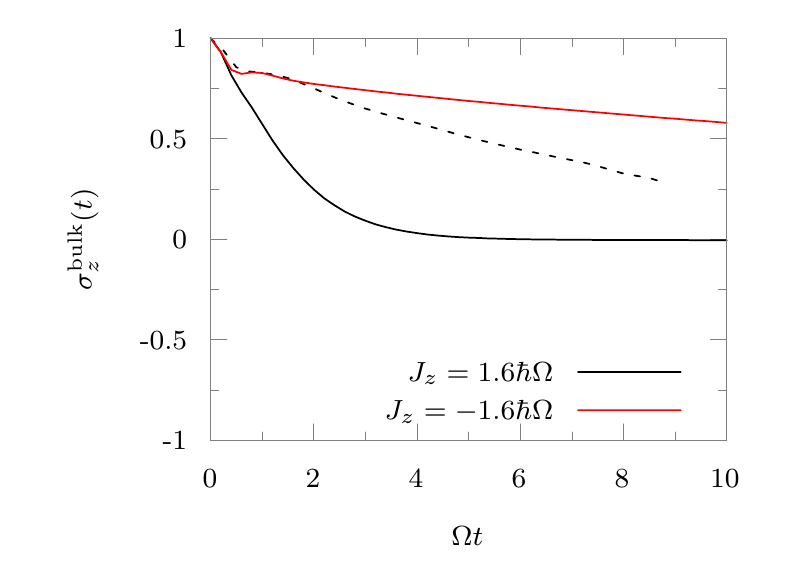}}

      \caption{Dynamics of a spin in the bulk as represented by
        $\expval{\hat{\sigma}_z(t)}$ for the 16\textsuperscript{th} site of the
        Ising model coupled to an Ohmic bath. Dashed line: without
        bath.}\label{fig:ising}
    \end{figure}
  \end{minipage}
\end{widetext}

\begin{figure}
  \includegraphics{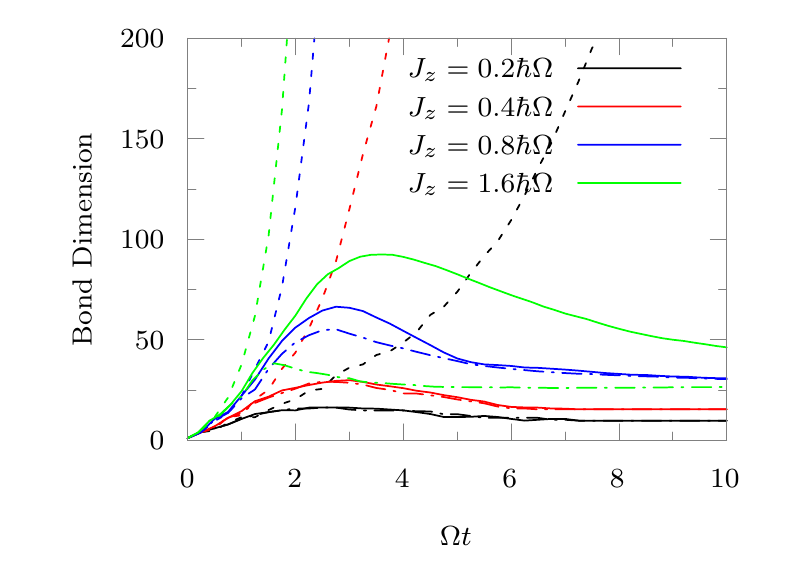}
  \caption{Average bond dimensions of the reduced density MPS for the bare
    (dashed line) and the full system for various intersite couplings. Positive
    $J_z$ is represented by the full solid line and negative $J_z$ by the
    dot-dashed line.}\label{fig:bonddims}
\end{figure}

There are two competing factors that contribute to the computational complexity
--- the non-Markovian memory caused by the presence of the bath, and the
entanglement that develops between the sites as time evolves causing the bond
dimension of the MPS to grow. The average bond dimension of the time evolved
density matrix can be used as a metric for the entanglement between the sites.
In Fig.~\ref{fig:bonddims}, we show its evolution as a function of time for the
parameters shown here. Though for the bare system case shown in
Fig.~\ref{fig:ising}, we propagated the wave function, here, for consistency,
the density matrix is propagated. We notice that the average bond dimension
grows faster for the bare system in comparison to all the cases with the bath,
demonstrating the decohering effect brought in by the dissipative medium. Though
the bath introduces a memory, it severely restricts the growth of the bond
dimension, and equivalently, the entaglement. It is interesting to note that for
higher values of interspin couplings, $J_z$, the difference in bond dimension
between the positive and the negative values increases. This is also reflected
in the fact that the dynamics becomes drastically different (cf.
Fig.~\ref{fig:ising}(c) and (d)). The eventual decrease in the bond dimension in
the presence of a bath reflects its ability to disentangle the system states.

\subsection{XXZ Model}

Another common model is the so-called XXZ-model, where the two-body interaction
term is defined by $J_x = J_y = J \ne 0$. In absence of any external field, the
ratio between $J_z$ and $J$ is an order parameter for quantum phase transitions
at zero temperature.~\cite{rakovSpinXXZHeisenberg10} When $J_z < -J$, the ground
state is ferromagnetic. There is a disordered spin-liquid phase when $-1 <
  \tfrac{J_z}{J} < 1$, and finally for $J_z > J$, there is an antiferromagnetic
state.

Here, we consider an XXZ system in an external transverse field of strength
$\Omega = 1$ and with a longitudinal field of $\epsilon = 0$. The harmonic bath
is once again held at an inverse temperature of $\hbar\Omega\beta = 1$. However,
in these examples, it is characterized by $\xi = 0.2$, $\omega_c = 2$. We
consider cases where $J_z = \pm 5J$ and $J_z = 0$. The time-step used for the
convergence is $\Omega\Delta t = 0.25$. The compression was done at a cutoff of
$\chi=10^{-11}$. The dynamics of $\expval{\hat{\sigma}_z(t)}$ for a bulk spin is
demonstrated for $J = 0.1$ and $J_x = 0,\pm0.5$ in Fig.~\ref{fig:XXZ}.

\begin{widetext}
  \begin{minipage}{\linewidth}
    \begin{figure}[H]
      \centering
      \subfloat[$J = 0.1$, $J_z = 0$.]{\includegraphics{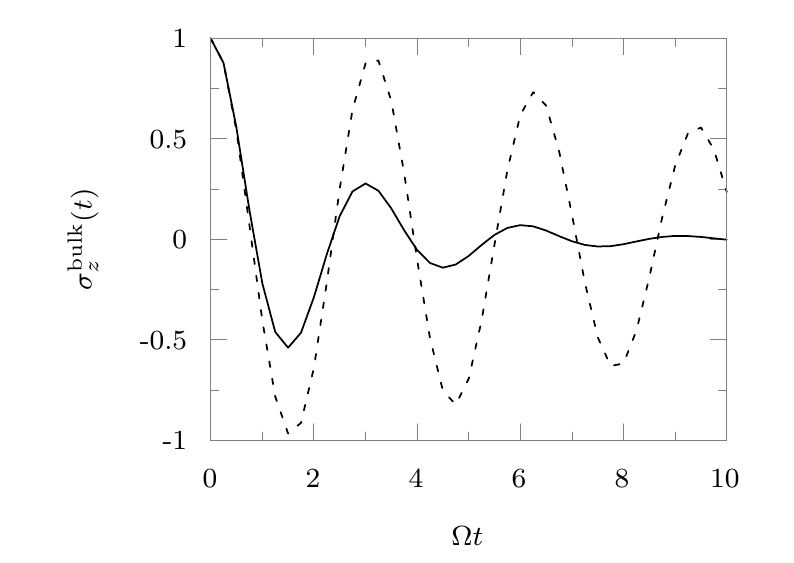}}
      ~\subfloat[$J = 0.1$, $|J_z| = 0.5$.]{\includegraphics{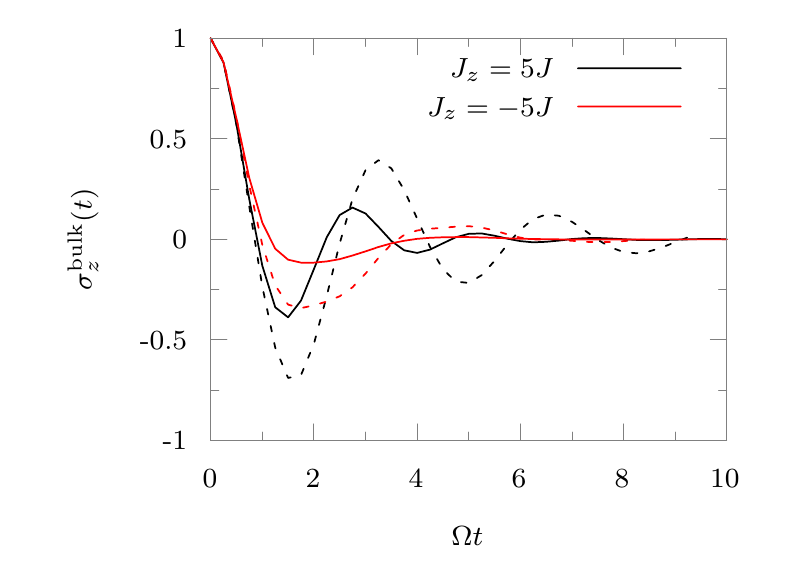}}
      \caption{Dynamics of a spin in the bulk as represented by
        $\expval{\hat{\sigma}_z(t)}$ for the 16\textsuperscript{th} site of the
        XXZ-model coupled to a harmonic bath. Dashed line: without
        bath.}\label{fig:XXZ}
    \end{figure}
  \end{minipage}
\end{widetext}

We note that the dynamics corresponding to the different values of $J_z$ are
totally different, even in the absence of the bath. It seems that the
dissipation effects increase as the absolute value of $J_z$ increases. However,
this increase in dissipation does not happen symmetrically. The effects coming
from a negative $J_z$ are much more pronounced than those caused by a positive
value. This difference may owe its origin to the fact that a negative $J_z$
stabilizes the initial state of all up spins, whereas positive values of $J_z$
destabilizes this state further. Also, the energy spectrum of the bare XXZ
system is different in the two cases. These differences and the full effects of
vibrational baths are very interesting and deserve a thorough analysis, that
will be the subject of future work.

\begin{figure}[h]
  \hspace*{-2cm}\includegraphics{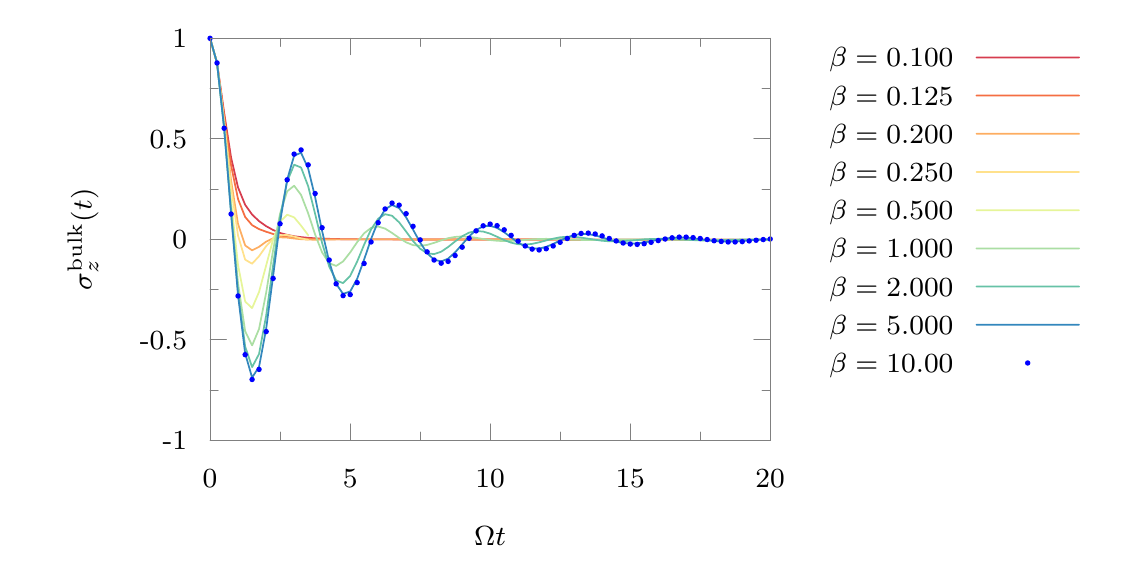}

  \caption{Dynamics of $\expval{\hat{\sigma}_z(t)}$ for the XXZ bulk spin (with
    $J_z = 0$) at different temperatures.}\label{fig:tempscan}
\end{figure}

Next, we study the effect of temperature on the dynamics of these XXZ systems.
  With wave-function methods like DMRG, it becomes especially difficult to do
  comparatively high temperature simulations because a large number of
  eigenstates of the bath needs to be accounted for. In Fig.~\ref{fig:tempscan},
  we demonstrate the dynamics of the $J_z = 0$ system at different temperatures
  ranging from $\beta = 10$ to $\beta = 0.1$. Temperature does not affect the
  converged value of $\Delta t$, but at higher temperatures the memory span
  needed to fully converge the dynamics increases. As per our intuitive
  understanding, at higher temperatures, the oscillations get washed away,
  showing a transition from an underdamped coherent oscillatory dynamics at low
  temperatures to a fully incoherent dynamics at high temperatures.

\subsection{Heisenberg Model}

The most general model for interacting spin chains is the so-called
``Heisenberg'' model. The Hamiltonian is characterized by a two-body spin-spin
interaction term that involves independent couplings along $X$, $Y$ and $Z$. The
bath used for this example is the same as the one used for the XXZ-model
examples.

\begin{figure}[h]
  \includegraphics{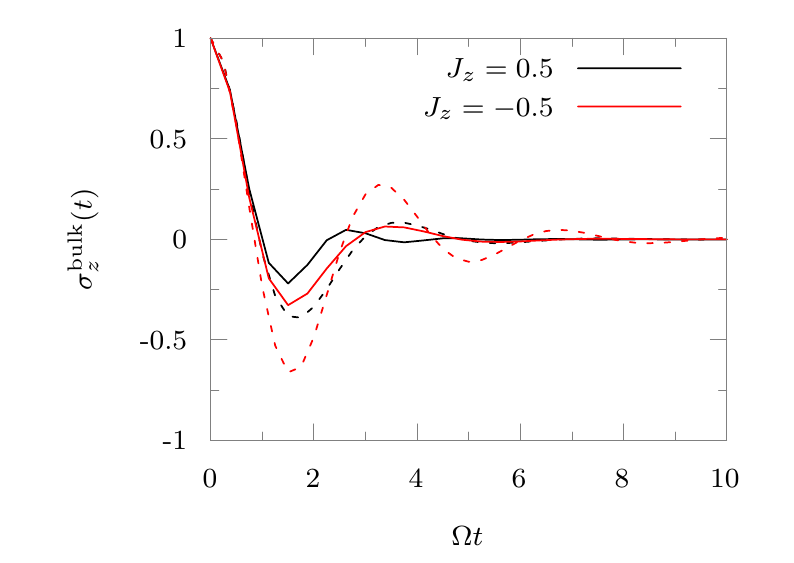}
  \caption{Dynamics of a spin in the bulk represented by
    $\expval{\hat{\sigma}_z(t)}$ for the 16\textsuperscript{th} site of the
    Heisenberg model coupled to a harmonic bath. Dashed line: without
    bath.}\label{fig:Heisen}
\end{figure}

The dynamics of $\expval{\hat{\sigma}_z(t)}$ for a state in the bulk is shown in
Fig.~\ref{fig:Heisen} for the case of $J_x = 0, J_y = 0.1, J_z = \pm0.5$. The
simulation was converged at a time-step of $\Omega\Delta t = 0.375$, a memory
length of $L = 3$ and a cutoff of $\chi=10^{-10}$. While we report the dynamics
of only the bulk spin here, for the case of $J_z = 0.5$, our simulations
reproduce the previously obtained results~\cite{kunduModularPathIntegral2020}
for the terminal edge spin.

\begin{figure}
  \includegraphics{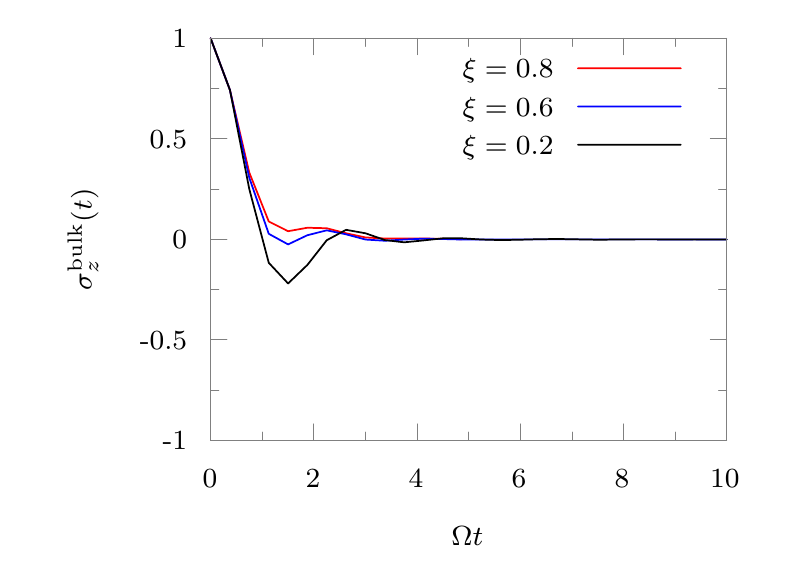}
  \caption{Dynamics of a bulk spin represented by $\expval{\hat{\sigma}_z(t)}$
    for the Heisenberg model with $J_z = 0.5$ and different values of Kondo parameters.}\label{fig:comp_coupl}
\end{figure}

Next, we explore the dissipation effects of the bath. In
  Fig.~\ref{fig:comp_coupl}, we demonstrate the dynamics of the bulk spin for
  the case of $J_z = 0.5$ where the bath is characterized by different Kondo
  parameters. The dynamics was converged at $L = 5$ for $\xi=0.6$ and $\xi =
  0.8$. The converged time-step $\Omega\Delta t$ remained unchanged at 0.375.
  One can see the oscillatory nature of the coherent dynamics decrease on
  increasing the coupling between the system and the bath due to the dissipative
  effects of the environment.

\section{Conclusion}\label{sec:conclusion}

System-solvent decomposition is used in various fields of study and is simulated
using the Feynman-Vernon influence functional. In such applications there is a
necessity to have a low-dimensional quantum system for computational
feasibility. However, many interesting problems involve extended systems that
can be modeled as collection of many low-dimensional units. Typical examples
involve spin-chains to model magnetism and charge or exciton transfers. Such
systems have historically been studied without any dissipative environment.
Early attempts at applying the same DMRG-based ideas led to interesting
applications that resorted to truncating a wave-function description of the
environment modes as well.~\cite{renTimeDependentDensityMatrix2018} MPI can
simulate such extended systems with an associated harmonic bath using
Feynman-Vernon influence functionals by iteratively incorporating the various
sites to side-step the exponential growth of storage and computational
requirements for extended quantum systems. However, it is often useful to
simulate the full density matrix, especially for many-body observables. In this
paper, we have developed a different approach to simulating these systems by
marrying tensor network structures with influence functionals.

Recently, various methods based on tensor networks have been proposed to
compress the path integrals, thereby reducing the storage requirements for the
simulation. Such compressions, if possible, for multisite systems would
especially be especially lucrative because the storage requirements generally
grow exponentially with the number of sites. Tensor network path integral
methods naturally suggest a further decomposition along the system dimension
leading to a multisite method. In this paper, we have introduced a multisite
tensor network framework called MS-TNPI, which solves the problem of an extended
quantum system coupled with dissipative environments extremely effectively. It
is a 2D extension of the 1D MPS structure used in the
AP-TNPI~\cite{boseTensorNetworkRepresentation2021} or the time-evolving matrix
product operators method
(TEMPO).~\cite{strathearnEfficientNonMarkovianQuantum2018}

The most essential part of MS-TNPI starts with the definition of a
forward-backward propagator for the extended system. This is a common problem
that is extensively dealt with in the
literature.~\cite{paeckelTimeevolutionMethodsMatrixproduct2019,renTimeDependentDensityMatrix2018,whiteRealTimeEvolutionUsing2004,daleyTimedependentDensitymatrixRenormalizationgroup2004}
We show how we can essentially use these propagators and refactorize them to
obtain the current 2D structure. We also discuss how by viewing the
aforementioned tensor network as a collection of rows containing the path
amplitude tensors corresponding to every site, it is now possible to apply the
influence functional MPO for the local bath in a systematic manner. Exploration
of other existing methods for MPO-MPS style wave function propagation in the
current context would be an interesting avenue of research. Especially methods
like W\textsuperscript{I,II} and TDVP, if adapted to the current framework,
would enable the simulation of significantly long-ranged interacting systems in
presence of local phononic modes. The decoupling of the structure of the system
Hamiltonian from the algorithm is an advantage of the current framework.

In general, the MS-TNPI structure can be used to calculate the AP for the
extended quantum system. In this paper, however, we have outlined efficient
algorithms that focus on using it to generate the full reduced density tensor
corresponding to the entire extended system at any point of time in the form of
an MPS. MS-TNPI, by its formulation, is also capable of handling arbitrarily
complex system initial states represented in the MPS form. While having this
global knowledge means that the storage requirements increase with the number of
sites, we show that the presence of the local vibrational bath helps decrease
the growth of the bond dimension of the reduced density MPS. This feature
enables trivial memory iterations corresponding to the finiteness of the
non-Markovian memory and efficient evaluation of many-body observables. While
the current development uses the analytical form for the influence functionals
derived for harmonic baths, it would be interesting to explore the prospects of
using the present structure with the more general numerical algorithm for
calculating the influence functionals as MPO derived by
\citeauthor{yeConstructingTensorNetwork2021}.\cite{yeConstructingTensorNetwork2021}

MS-TNPI is demonstrated through illustrative examples of various spin-chains
coupled to local harmonic baths. We simulate the Ising model, the XXZ-model and
the Heisenberg model with various different parameters. We show that the
site-local baths severely restrict the growth of the bond dimension in the
reduced density MPS. Consequently, in comparison to the bare system, the
intersite entanglement grows slower and often even decreases in the presence of
dissipative environments. We have explored the three different causes of
decohering of the system dynamics. There are interesting features of the
dynamics for the various phases of the XXZ-model that require further
investigation. This would be the subject of future work.

MS-TNPI promises to be an exciting method for extended systems. It makes it
possible to study various energy and charge transfer processes, and loss of
coherence in chains of qubits. Such applications shall be the focus of our
research in the near future. Additionally, the novelty of the structure opens up
possibilities for further improvements and developments of which we have only
begun scratching the surface.

\section*{Authors' Contributions}
Both authors contributed equally to this work.

\section*{Acknowledgments}
A.\,B. acknowledges the support of the Computational Chemical Center: Chemistry
in Solution and at Interfaces funded by the US Department of Energy under Award
No. DE-SC0019394. P.\,W. acknowledges the Miller Institute for Basic Research in
Science for funding.

\section*{Data Availability}
The data that support the findings of this study are available from the
corresponding author upon reasonable request.

\appendix
\section{System Forward-Backward Propagator in MPO
  representation}\label{app:MPOprop}

For the simple case of nearest-neighbor interacting Hamiltonian, it is very easy
to define an algorithm for calculating the second-order Suzuki-Trotter split
forward-backward propagator. This is a ``forward-backward'' version of the
second-order time-evolved block decimation
scheme.~\cite{paeckelTimeevolutionMethodsMatrixproduct2019}

Let the Hamiltonian be factorized as:
\begin{align}
  \hat{H}_0 = \sum_{i=1}^{P-1} \hat{\mathcal{H}}_{i,(i+1)}
\end{align}
where $\hat{\mathcal{H}}_{i,(i+1)}$ takes the one-body term into account as well. To get the
second-order splitting, it is usual to incorporate the terminal single body
terms fully into the corresponding terms. Everything else is split in halves:
\begin{align}
  \hat{\mathcal{H}}_{1,2}     & = \hat{h}^{(1)}_1 + \hat{h}^{(2)}_{1,2} + \frac{1}{2}\hat{h}^{(1)}_2                                       \\
  \hat{\mathcal{H}}_{i,(i+1)} & = \frac{1}{2}\hat{h}^{(1)}_i + \hat{h}^{(2)}_{i,(i+1)} + \frac{1}{2}\hat{h}^{(1)}_{(i+1)},\quad 2\le i<P-1 \\
  \hat{\mathcal{H}}_{(P-1),P} & = \frac{1}{2}\hat{h}^{(1)}_{(P-1)} + \hat{h}^{(2)}_{(P-1),P} + \hat{h}^{(1)}_P.
\end{align}

Traditionally, these terms are grouped as ``even'' and ``odd'' as follows:
  \begin{align}
    \hat{\mathcal{H}}_{\text{odd}}  & = \sum_{1\le i<P}^{i\text{ odd}}  \hat{\mathcal{H}}_{i,(i+1)}  \\
    \hat{\mathcal{H}}_{\text{even}} & = \sum_{1\le i<P}^{i\text{ even}} \hat{\mathcal{H}}_{i,(i+1)},
  \end{align}
  where $\hat{\mathcal{H}}_{\text{odd}}$ and $\hat{\mathcal{H}}_{\text{even}}$ do
  not commute. Under the second order Suzuki-Trotter factorization the
  forward-backward propagator,
  \begin{align}
    K\left(S^\pm_{n}, S^\pm_{n+1}, \Delta t\right) & \approx \sum_{S^\pm_{n^\prime}, S^\pm_{n^{\prime \prime}}} K_{\text{odd}}(S^\pm_{n}, S^\pm_{n^\prime}, \tfrac{1}{2}\Delta t)\nonumber \\
                                                   & \times K_{\text{even}}(S^\pm_{n^\prime}, S^\pm_{n^{\prime \prime}}, \Delta t)\nonumber                                                \\
                                                   & \times K_{\text{odd}}(S^\pm_{n^{\prime \prime}}, S^\pm_{n+1}, \tfrac{1}{2}\Delta t),\label{eq:prop_TEBD}
  \end{align}
  where $S^\pm_{n^\prime},$ and $S^\pm_{n^{\prime \prime}}$ are dummy variables
  that represent the forward-backward state of the system at the two intermediate
  time points. In the notation used here, $K_{\text{odd}}(S^\pm_{n},
    S^\pm_{n+1},\Delta t)$ and $K_{\text{even}}(S^\pm_{n}, S^\pm_{n+1},\Delta t)$
  are the forward-backward propagators associated with
  $\hat{\mathcal{H}}_{\text{odd}}$ and $\hat{\mathcal{H}}_{\text{even}}$,
  respectively. The odd and even terms commute with themselves, so the
  forward-backward propagators can be factorized further:
  \begin{align}
    K_{\text{odd}}\left(S^\pm_{n}, S^\pm_{n+1}, \Delta t\right)  = & \prod_{1\le i<P}^{i\text{ odd}}\nonumber  \\ &K_{i}(s^\pm_{i,n}, s^\pm_{i+1,n}, s^\pm_{i,n+1}, s^\pm_{i+1,n+1},\Delta t)\label{eq:FB_Prop_odd} \\
    K_{\text{even}}\left(S^\pm_{n}, S^\pm_{n+1}, \Delta t\right) = & \prod_{1\le i<P}^{i\text{ even}}\nonumber \\ &K_{i}(s^\pm_{i,n}, s^\pm_{i+1,n}, s^\pm_{i,n+1}, s^\pm_{i+1,n+1},\Delta t),\label{eq:FB_Prop_even}
  \end{align}
  where the two-body forward-backward propagator,
  \begin{align}
    K_{i}(s^\pm_{i,n} & , s^\pm_{i+1,n}, s^\pm_{i,n+1}, s^\pm_{i+1,n+1},\Delta t) =\nonumber                                                                      \\
                      & \mel{s^+_{i,n+1}, s^+_{i+1,n+1}}{\exp\left(-\frac{i}{\hbar} \hat{\mathcal{H}}_{i,(i+1)} \Delta t\right)}{s^+_{i,n}, s^+_{i+1,n}}\nonumber \\
                      & \times \mel{s^-_{i,n}, s^-_{i+1,n}}{\exp\left(\frac{i}{\hbar} \hat{\mathcal{H}}_{i,(i+1)} \Delta t\right)}{s^-_{i,n+1}, s^-_{i+1,n+1}}.
  \end{align}

Using an SVD, we can factor this two-body (multi-site) operator into a pair of
single-body (single-site) operators:
  \begin{align}
    K_{i}(s^\pm_{i,n}, & s^\pm_{i+1,n}, s^\pm_{i,n+1}, s^\pm_{i+1,n+1},\Delta t) =\nonumber                                                                            \\
                       & \sum_{\alpha_{(i,n)}} W^{s^\pm_{i,n}, s^\pm_{i,n+1}}_{\alpha_{(i,n)}} W^{s^\pm_{i+1,n}, s^\pm_{i+1,n+1}}_{\alpha_{(i,n)}}.\label{eq:pair-MPO}
  \end{align}
  Plugging this expression into Eqs.~\ref{eq:FB_Prop_odd} and ~\ref{eq:FB_Prop_even} gives:
  \begin{align}
    K_{\text{odd}}\left(S^\pm_{n}, S^\pm_{n+1}, \Delta t\right)  = & \prod_{1\le i<P}^{i\text{ odd}} \left(\sum_{\alpha_{(i,n)}} W^{s^\pm_{i,n}, s^\pm_{i,n+1}}_{\alpha_{(i,n)}} W^{s^\pm_{i+1,n}, s^\pm_{i+1,n+1}}_{\alpha_{(i,n)}}\right)\label{eq:PropMPO_odd}    \\
    K_{\text{even}}\left(S^\pm_{n}, S^\pm_{n+1}, \Delta t\right) = & \prod_{1\le i<P}^{i\text{ even}} \left(\sum_{\alpha_{(i,n)}} W^{s^\pm_{i,n}, s^\pm_{i,n+1}}_{\alpha_{(i,n)}} W^{s^\pm_{i+1,n}, s^\pm_{i+1,n+1}}_{\alpha_{(i,n)}}\right)\label{eq:PropMPO_even}.
  \end{align}
  Notice that by introducing unit-dimensional bond indices between the products,
   they can be reduced to proper MPOs. Thus, the resulting MPOs would have
   alternating bonds of unit dimension. Finally, the full second order
   Suzuki-Trotter split forward-backward propagator can be obtained through
   sequential MPO-MPO multiplications involving Eqs.~\ref{eq:PropMPO_odd} and
   \ref{eq:PropMPO_even} as per Eq.~\ref{eq:prop_TEBD}.

\bibliography{library}
\end{document}